\documentclass[preprint,journal,twoside,web]{ieeecolor}
\usepackage{tbme}
\usepackage{cite}
\usepackage{amsmath,amssymb,amsfonts}
\usepackage{graphicx}
\usepackage{textcomp}
\usepackage{url}
\usepackage{algorithm,algcompatible,amsmath}
\usepackage{algorithmicx}
\usepackage{amssymb}
\makeatletter
\newcommand*{\toccontents}{\@starttoc{toc}}
\makeatother
\usepackage{comment}
\newcommand{\C}{\mathbb C}
\newcommand{\R}{\mathbb R}

\newcommand{\XX}{\mathbf{x}} 
\newcommand{\PP}{\mathbf{p}} 
\newcommand{\UU}{\mathbf{u}} 
                       
\newcommand{\zz}{\mathbf{z}}                     
\newcommand{\YY}{\mathbf{y}}

\newcommand{\Ad}{\mathbf A}         
\newcommand{\Zd}{\mathbf Z}

\newcommand{\Id}{\mathbf I}                        
\newcommand{\Cd}{\mathbf C}                        
\newcommand{\Ed}{\mathbf E}                        
\newcommand{\Hd}{\mathbf H}

\newcommand{\Wd}{\mathbf W}
\newcommand{\Pd}{\mathbf P}

\newcommand{\Au}{\mathbf{A}_I}

\newcommand{\Yu}{\mathbf{y}_I}          
                     
\newcommand{\herm}{{\scriptstyle \boldsymbol{\mathsf{H}}}}
\newcommand{\trans}{{\scriptstyle \boldsymbol{\mathsf{T}}}}

\def\BibTeX{{\rm B\kern-.05em{\sc i\kern-.025em b}\kern-.08em
    T\kern-.1667em\lower.7ex\hbox{E}\kern-.125emX}}
\usepackage{afterpage}

\newcommand\blankpage{
    \null
    \thispagestyle{empty}
    \newpage
    }

\begin{document}

\afterpage{\blankpage}
\begin{center}
    \begin{minipage}{18cm}
    \centering
    \Large{Quantitative MR Image Reconstruction using Parameter-Specific Dictionary Learning with Adaptive Dictionary-Size and Sparsity-Level Choice}\\ \vspace{0.2cm}
    \large{A. Kofler, K. M. Kerkering, L. Göschel, A. Fillmer, C. Kolbitsch.}\\ \vspace{0.8cm}

    \raggedright{
    $\copyright$ 2023 IEEE. This paper has been accepted for publication in IEEE Transactions on Biomedical Engineering.
    This is the author’s version of an article that has, or will be, published in this journal or conference.
    Changes were, or will be, made to this version by the publisher prior to publication.\\ \vspace{0.2cm}
    
    DOI: 10.1109/TBME.2023.3300090 \\
    IEEE Xplore: https://ieeexplore.ieee.org/document/10209588\\ \vspace{0.5cm}
    
    Please cite this paper as:\\
    A. Kofler, K. M. Kerkering, L. Göschel, A. Fillmer, C. Kolbitsch. “Quantitative MR Image Reconstruction using Parameter-Specific Dictionary Learning with Adaptive Dictionary-Size and Sparsity-Level Choice,” in IEEE Transactions on
    Biomedical Engineering, doi: 10.1109/TBME.2023.3300090. \\ \vspace{0.5cm}

    \textbf{Acknowledgments}: The results presented here have been developed in the
framework of the 18HLT05 QUIERO Project and 18HLT09 NeuroMET2 Project. These projects have received
funding from the EMPIR programme co-financed by the
participating states and from the European Union’s Horizon
2020 research and innovation program.
}
    \end{minipage}
\end{center}

\newpage

\title{Quantitative MR Image Reconstruction using Parameter-Specific Dictionary Learning with Adaptive Dictionary-Size and Sparsity-Level Choice}
\author{Andreas Kofler, Kirsten Miriam Kerkering, Laura Göschel, Ariane Fillmer, Christoph Kolbitsch
\thanks{Date of Submission: August 11, 2022. 
The results presented here have been developed in the framework of the 18HLT05 QUIERO project. Further, the project 18HLT09 – NeuroMET2 has received funding from the EMPIR program co- financed by the Participating States and from the European Union’s Horizon 2020 research and innovation program.}
\thanks{A. Kofler, K. Becker, A. Fillmer and Christoph Kolbitsch are with the 
Physikalisch-Technische Bundesanstalt, Braunschweig and Berlin, Germany (e-mails: $\{$andreas.kofler, kirsten.kerkering, ariane.fillmer, christoph.kolbitsch$\}$@ptb.de). L. Göschel is with the Charité - Universitätsmedizin Berlin, Berlin, Germany (e-mail: laura.goeschel@charite.de).  }
}

\maketitle

\begin{abstract}
Objective: We propose a method for the reconstruction of parameter-maps in Quantitative Magnetic Resonance Imaging (QMRI).\\
Methods: Because different quantitative parameter-maps differ from each other in terms of local features, we propose a method where the employed dictionary learning (DL) and sparse coding (SC) algorithms automatically estimate the optimal dictionary-size and sparsity level separately for each parameter-map. We evaluated the method on a $T_1$-mapping QMRI problem in the brain using the BrainWeb data as well as in-vivo brain images acquired on an ultra-high field 7\,T scanner. We compared it to a model-based acceleration for parameter mapping (MAP) approach, other sparsity-based methods using total variation (TV), Wavelets (Wl) and Shearlets (Sh), and to a method which uses DL and SC to reconstruct qualitative images, followed by a non-linear (DL+Fit).\\
Results: Our algorithm surpasses MAP, TV, Wl and Sh in terms of RMSE and PSNR. It yields better or comparable results to DL+Fit by additionally significantly accelerating the reconstruction by a factor of approximately seven.\\
Conclusion: The proposed method outperforms the reported methods of comparison and yields accurate $T_1$-maps. Although presented for $T_1$-mapping in the brain, our method’s structure is general and thus most probably also applicable for the the reconstruction of other quantitative parameters in other organs.\\
Significance: From a clinical perspective, the obtained $T_1$-maps could be utilized to differentiate between healthy subjects and patients with Alzheimer's disease. From a technical perspective, the proposed unsupervised method could be employed to obtain ground-truth data for the development of data-driven methods based on supervised learning.
\end{abstract}

\begin{IEEEkeywords}
Dictionary Learning, Quantitative Imaging, $T_1$-mapping, Compressed Sensing
\end{IEEEkeywords}
\newpage
\section{Introduction}
\label{sec:introduction}
\IEEEPARstart{M}{agnetic} Resonance Imaging (MRI) is a nowadays indispensable tool in every day's clinical routine. Most commonly, only qualitative images are reconstructed in which the contrast between healthy and diseased tissue is used for diagnostic purposes. The main challenge of qualitative imaging is that the image intensities do not just depend on the underlying tissue but also on the MR sequence used for the data acquisition as well as on hardware-specific factors. \\
Quantitative MRI (QMRI) overcomes this challenge by directly providing (bio-) physical parameters and thus facilitating the  comparison among different measurements. Especially some neurological applications where the differences in contrast between healthy and diseased tissue are subtle, such as the early diagnosis of neurodegenerative diseases like Alzheimer's disease, quantitative imaging of physical parameters such as $T_1$ relaxation might be a game changer. 
It is known that the neuronal density as well as the macromolecular content of brain tissue change in neurodegenerative diseases due to the accumulation of protein plaques and tangles. These changes in the tissue microstructure are expected to cause changes of the relaxation behaviour of the affected brain tissues. Nevertheless, the literature on the usefulness of $T_1$ relaxation changes in Alzheimer's disease is sparse and inconclusive, see \cite{tang2018magnetic} for a concise literature review. This inconclusiveness seems to be mostly attributable to the use of rather low field strengths in studies so far as well as to high measurement uncertainties due to the long acquisition times for individual time points required for the subsequent $T_1$-fit.\\
Thus, it is desirable to accelerate the measurement process as well as to increase the field strength. However, in order to ensure accurate parameter estimation even from short scan times, i.e.\ from highly undersampled $k$-space data, regularized image reconstruction approaches are required.\\
In recent years, several approaches have been proposed which combine model-based parameter reconstruction with sparsity-based regularization methods, e.g.\ total variation (TV)-minimization \cite{Zhao2014, Wang2019b} as well as Wavelets \cite{Hamilton2017} or Shearlets  \cite{Ma2017c}. Dictionary learning (DL) is a well-established data-driven regularization method which has been extensively applied for qualitative MR-image reconstruction, see e.g. \cite{ravishankar2011mr, caballero2014dictionary, wang2014compressed, song2019coupled, pali2021adaptive}. Mathematically, a dictionary is referred to as an overcomplete basis which can be  used to approximate signals of interest using only a small number of basis functions $-$ the so-called atoms. By the lower-dimensional approximation, noise and artefacts are discarded and only the most relevant information about the signal is kept, yielding a powerful regularization method. However, identifying the atoms which yield the best sparse approximation, i.e.\ the sparse coding (SC) of the signal, is a challenging (and typically time-consuming) problem which has been extensively studied in the signal processing community. Further, the number of overall available atoms $K$ as well as the maximal number of possibly used atoms $S$ per signal has to be carefully chosen in advance. Overestimating $K$ as well as $S$ can  lead to significantly longer reconstruction times and at the same time yield worse reconstruction results because of overfitting. To counter that, $K$ and $S$-adaptive DL and SC methods in which the dictionary size and/or the sparsity level are estimated based on the considered data were considered in the signal processing community \cite{do2008sparsity}, \cite{blumensath2009stagewise}, \cite{donoho2012sparse}, \cite{wu2012adaptive}, \cite{pali2018dictionary} and successfully applied to (qualitative) MRI reconstruction problems \cite{ahishakiye2020adaptive}, \cite{pali2021adaptive}.\\
In \cite{pali2021adaptive}, the authors demonstrated that using the $K$-$S$-adaptive DL and SC algorithms adaptive iterative thresholding and $K$ residual means (aITKrM) \cite{pali2018dictionary} and adaptive orthogonal matching pursuit (aOMP) \cite{pali2021adaptive}, reconstruction times could be reduced while maintaining the same reconstruction quality compared to using $K$-SVD \cite{aharon2006k} and OMP \cite{pati1993orthogonal} which are the typical methods of choice for DL and SC.\\
For the task of quantitative parameter mapping, however, DL and SC are by far not as popular as for qualitative MRI. Typically, when used for QMRI applications, DL and SC are used to first reconstruct a set of qualitative images from which, in a second step, quantitative parameter-maps are estimated using a non-linear fit, see e.g.\ \cite{Doneva2010}, \cite{zhu2019integrated}. The computational bottleneck of these approaches is  the relatively long reconstruction time required to reconstruct all qualitative images for each sampling point before the non-linear optimization routine can can be run to obtain the quantitative parameters \cite{tamir2020computational}. In this work, we circumvent this issue by directly imposing the DL- and SC-regularization  on the quantitative parameters.  By doing so, the SC-step $-$ which is the computationally most expensive part $-$  has only to be performed with respect to the quantitative parameter-maps instead as for all intermediate qualitative images. A similar approach has been for example investigated in \cite{Zhao2014} for fixed and non-learned orthonormal sparsifying transforms, e.g. Wavelets. However, using pre-defined sparsifying transforms might be sub-optimal. For example, Wavelets are known to suffer from blocking and smoothing artefacts \cite{lustig2008compressed}. In this work, we therefore use DL and SC as local patch-wise transformations learned from data. Further, to count for the fact that quantitative parameter maps can highly vary from each other in terms of local features, we adopt the aforementioned adaptive DL and SC algorithms such that each dictionary is tailored to its respective quantitative parameter. We evaluate our method on a $T_1$-mapping image reconstruction problem in the brain and compare it to TV-, Wavelet- and Shearlet-based methods as well as to a DL- and SC-based approach which imposes the regularization on the intermediate qualitative images. We show that the proposed approach consistently outperforms the non-ML methods and yields better or comparable reconstructions to DL and SC when applied to regularize the qualitative images while only requiring about a seventh of the reconstruction time.\\
Although the proposed method is presented for a brain QMRI reconstruction problem using DL and SC, we expect the employed reconstruction strategy based on splitting the linear measurement model from the non-linear signal model to be also applicable for other QMRI problems as well as for other potentially time-consuming regularization methods which involve an optimization problem, such as convolutional DL \cite{chun2017convolutional}, \cite{garcia2018convolutional}, or convolutional analysis operator learning \cite{Chun2020ConvolutionalAO}.\\
The paper is organized as follows. In Section \ref{sec:problem_formulation}, we introduce the notation used throughout the manuscript and formally state the reconstruction problem. In Section \ref{sec:proposed_method}, we motivate and introduce the proposed reconstruction method for obtaining the quantitative parameters from the undersampled $k$-space data. In Section \ref{sec:experiments}, we introduce the  signal-model which was used for the evaluation of the proposed method as well as the methods of comparison. In Section \ref{sec:results} we show qualitative and quantitative results obtained with the reported methods. In Section \ref{sec:discussion}, we discuss different components of the proposed method and highlight similarities and differences as well as advantages and limitations of the proposed method and conclude the work with Section \eqref{sec:conclusion}.

\section{Problem Formulation}\label{sec:problem_formulation}
In the following, we introduce the notation used throughout the manuscript needed to describe the model and the operators required for the  formulation of the reconstruction problem  we aim to solve.
\subsection{The Signal Model}
We denote by $\PP$ the vector representation of the (unknown) quantitative parameter-maps which we are interested in. Depending on the application, the vector $\PP$ can contain $P$ different quantities, each given by an $N$-dimensional real- or also complex-valued vector. For example, $\PP$ could refer to different relaxation times, i.e. $\PP=[\mathbf{T}_1, \mathbf{T}_2, \mathbf{T}_2^\ast]^\trans \in \mathbb{R}^{3N}$ or to physiological parameters such as susceptibility- or diffusion-related parameters. 
Let $V \in \{ \R, \C\}$. For a time point $t>0$ , we define the signal model $q_t$ by the mapping 
\begin{eqnarray}\label{eq:q_t}
q_t: &V^{PN} &\rightarrow \C^N, \\
&\PP=[\PP_1,\ldots,\PP_P]^\trans &\mapsto q_t(\PP),
\end{eqnarray}
  which describes the interaction of the different quantities contained in $\PP$. Then, for a set of time points $\mathcal{T}=\{t_1, \ldots, t_T\}$ with $t_{i} < t_{i+1}$, we define $q_{\mathcal{T}}$ by
\begin{eqnarray}\label{eq:q}
q_{\mathcal{T}}: & V^{PN} \rightarrow &\C^{TN}, \\
&\PP \mapsto &q_{\mathcal{T}}(\PP) = [q_{t_1}(\PP), \ldots, q_{t_T}(\PP)]^\trans
\end{eqnarray}
and identify $q_{\mathcal{T}}$ with the process $(q_t)_{t\in \mathcal{T}}$.

\subsection{The Measurement Model}
The measurement process then takes place in the Fourier-domain of the vector $q_{\mathcal{T}}(\PP)$. At the different time points indexed by $\mathcal{T}$ an operator (which assuming a homogeneous static magnetic field can be described by a Fourier-transform) samples the vector $q_t(\PP)$ for each $t\in \mathcal{T}$. Recall that $q_{\mathcal{T}}(\PP) \in \C^{TN}$ contains information about the interaction of the different quantities in $\PP$ at all different time points $t_i$, $1<i<T$. In addition, it is desirable to accelerate the acquisition time for each time point by undersampling the quantity $q_{t_i}(\PP)$. In order to collect complementary information about the entire process $(q_t(\PP))_{t\in \mathcal{T}}$,  the sampling trajectories can be chosen to differ among the acquisition points $t_i, i=1, \ldots, T$.\\
In addition, in clinical applications, it is common practice to employ multiple receiver coils for the acquisition of the data. By $\Cd = [\Cd_1, \ldots, \Cd_{N_c}]^\trans \in \mathbb{C}^{N_c\cdot N \times N}$ we denote a tall operator with $\Cd_i = \mathrm{diag}(\mathbf{c}_i)$, $\mathbf{c}_j \in \C^N$ containing the entries of the $i$-th coil-sensitivity map. Let $J$ denote the set of indices of the coefficients in $k$-space which need to be acquired in order to sample an image $\XX \in \C^N$ with $N=N_x \times N_y$ at Nyquist-limit and let $I_t \subset J$ denote a subset of $J$. Then, for a  single time-point $t>0$, the operator $\Ad_{I_t}$ is given by
\begin{equation}
    \Ad_{I_t}:= (\Id_{N_c} \otimes \Ed_{I_t}) \Cd,
\end{equation}
where $\otimes$ denotes the Kronecker-product, $\Id_{N_c}$ is an identity operator and the encoding operator $\Ed_{I_t}$ denotes a (possibly non-uniform) Fourier-encoding operator which samples the image $q_t(\PP)$ along the $k$-space trajectories implicitly defined by $I_t$. 
Finally, our considered forward problem is given by
\begin{equation}\label{eq:forward_problem}
\YY = \Au q_{\mathcal{T}}( \PP) + \mathbf{e},
\end{equation}
where $\Au:V^{TN} \rightarrow \mathbb{C}^M$ with $\Au :=  \mathrm{diag}(\Ad_{I_{t_1}},\ldots,\Ad_{I_{t_T}})$ and $I = \cup_{t \in \mathcal{T}} I_t$ is a block-diagonal operator and $\mathbf{e} \in \C ^M$ with $M= N_c \cdot (|I_{t_1}| + \ldots + |I_{t_T}|) $ denotes Gaussian noise, where $|I_{t_i}|$ is the cardinality of the index set $I_{t_i}$. From \eqref{eq:forward_problem}, we see that the operator $\Au  q_{\mathcal{T}} : V^{PN} \rightarrow \C^M$ maps a real- or complex-valued vector  $\PP=[\PP_1, \ldots, \PP_P]^\trans$ to an $M$-dimensional  complex-valued  measurement-vector $\YY$. \\
The goal is to recover the quantitative parameters $\PP$ from the measurements $\YY$. The realistic set-up is the case $M\gg N$ such that the recovery from the measurement vector $\YY$ is possible.\\
Due to the structure of the operator $\Au  q_{\mathcal{T}}$, it is possible to identify two aspects which make the reconstruction challenging.\\
First, for each time point $t\in \mathcal{T}$, the operator $\Ed_{I_t}$ under-samples the quantity  $q_t(\PP)$ by violating the Nyquist-limit in order to accelerate the scan. In general, the use of multiple receiver coils in parallel imaging, which collect complementary information, counters this undersampling. Algebraically speaking, the use of multiple coils changes the underlying problem from an underdetermined to an overdetermined one, making the inversion process possible in principle but ill-conditioned in practice, see e.g.\cite{qu2005convergence}. In addition, the application of the operator $\Au$ is possibly time-consuming and a repeated evaluation of the latter, as for example in non-linear methods which involve line-searches to determine step-sizes, must be avoided. Second, in general, the signal model $q_{\mathcal{T}}$ is non-linear, making the reconstruction process from noisy data even more challenging. As a consequence, the use of efficient regularization methods is essential for the recovery of the vector $\PP$.\\
In this work, we focus our approach on the use of patient-adaptive data-driven methods, where the regularization method itself is learned within the reconstruction process. Further, as we shall see later, our problem formulation aims at substantially reducing the time required for learning the regularization for possibly time consuming methods as DL/convolutional DL-based methods.

\section{Proposed Method}\label{sec:proposed_method}
Here, we briefly revise DL and SC  before describing its application in the proposed approach for quantitative MR image reconstruction.

\subsection{Dictionary Learning and Sparse Coding}\label{subsec:dico_learning}
Typically, a dictionary $\mathbf{\Psi}$ is defined to be a collection of $K$ $d$-dimensional basis functions $-$ so-called atoms $-$ with unit norm, i.e.\ $\mathbf{\Psi} = [\boldsymbol{\psi}_1, \ldots, \boldsymbol{\psi}_K ]$ with $\boldsymbol{\psi}_k \in \mathbb{R}^d$ and $\| \boldsymbol{\psi}_k \|_2 = 1$.
In DL, the goal is learn to decompose a data matrix $\Zd = [\zz_1, \ldots, \zz_L]$ with $\zz_k \in \mathbb{R}^d$ into a dictionary $\mathbf{\Psi} \in \mathbb{R}^{d\times K}$ with $d\leq K$ and a column-wise sparse coefficient matrix $\boldsymbol{\Gamma} = [\boldsymbol{\gamma}_1,\ldots, \boldsymbol{\gamma}_L] \in \mathbb{R}^{K \times L}$, i.e. each signal $\zz_k$ is represented by a linear combination of at most $S$ of the $K$ atoms of the dictionary $\mathbf{\Psi}$. The reason to constrain the atoms to have unit-norm is to avoid the scaling ambiguity between the dictionary and the sparse codes \cite{gribonval2010dictionary}. To measure the sparsity of a vector, the $\ell_0$-pseudo-norm is used, where $\| \cdot \|_0$ simply counts the non-zero entries of the vector. For a pre-defined sparsity level $S$, the DL problem is then given by
\begin{align}\label{DL_problem_L0}
\underset{\mathbf{\Psi} \in \mathcal{D}_{d,K}, \, \{\boldsymbol{\gamma}_j\}_j}{\mathrm{min}} \frac{1}{2}\sum_{j=1}^L \| \zz_j - \mathbf{\Psi} \boldsymbol{\gamma}_j\|_2^2 \quad \text{s.t.}\quad  \forall j: \| \boldsymbol{\gamma}_j\|_0 \leq S,
\end{align}
where $\mathcal{D}_{d,K}:=\{ \mathbf{\Psi} \in \R^{d\times K}: \forall k: \| \mathbf{\psi}_k \|_2 = 1 \}$.
Typically, the solution of problem \eqref{DL_problem_L0} is approached by means of alternating minimization, i.e.\ one  optimizes the dictionary $\mathbf{\Psi}$ assuming a fixed set of sparse codes $\{\boldsymbol{\gamma}_j\}_j$ and then optimizes the set of sparse codes by assuming a fixed dictionary. 

\subsection{Adaptive Iterative Thresholding and $K$ Residual Means and Adaptive Orthogonal Matching Pursuit}
Before proceeding with the presentation of the proposed method, we briefly revise the  basic ideas of the employed adaptive DL- and SC-algorithms used in this work. 
One of the most popular DL algorithms for solving \eqref{DL_problem_L0}  is the well-known $K$-SVD method \cite{aharon2006k} which alternates between computing a singular value decomposition (SVD) to update each atom and orthogonal matching pursuit (OMP) \cite{pati1993orthogonal} to sparsely approximate the signals. In this work, however, we use a sparsity-level and dictionary-size adaptive DL method $-$ adaptive iterative thresholding and $K$ residual means (aITKrM) \cite{pali2018dictionary} with a sparsity-adaptive OMP $-$ adaptive OMP (aOMP) for DL and SC \cite{pali2021adaptive}, respectively.
In \cite{pali2021adaptive}, the advantages of the combination aITKrM + aOMP compared to $K$-SVD + OMP were investigated for a non-Cartesian dynamic MR image reconstruction problem. 
There, the authors reported similar or improved results compared to $K$-SVD + OMP as well as a significant acceleration in terms of reconstruction times.\\ 
The first reason for the acceleration of the DL component by factor of approximately 10 obtained in \cite{pali2021adaptive} is the choice of aITKrM (which uses ITKrM) over $K$-SVD. The second reason is the use of aOMP instead of OMP. More specifically, the higher $S$ is chosen, the longer the sparse approximation takes and the times required for the SC of all image patches $-$ which is the computationally most demanding component in DL+SC-based image reconstruction $-$ was observed to be reduced by a factor of approximately five \cite{pali2021adaptive}. Last, from a practical point of view, the major advantage of aITKrM + aOMP compared to $K$-SVD + OMP is the fact that no hyper-parameter tuning of $S$ and $K$ is required. Instead, $S$ and $K$ are adaptively chosen depending on the data under consideration, i.e.\ based on the image patches of the current image estimate.\\
More precisely, for the DL stage, aITKrM involves strategies to replace too similar or rarely used atoms in the dictionary by promising atom candidates. These strategies stem from an analysis of the contractive behaviour of ITKrM and were first proposed in \cite{pali2023dictionary}. Further, the sparsity level during the DL stage, i.e.\ using iterative thresholding as SC algorithm, is also adaptively varied.\\
Last, when finally solving problem \eqref{DL_problem_L0} with respect to the sparse codes $\{\boldsymbol{\gamma}_j\}_j$, aOMP replaces the stopping criterion of OMP by introducing a bound for the maximal inner product between atoms and the current residual. Before including an atom to be used for  the sparse representation, aOMP checks if there exists an atom in the dictionary which is worthy to be used. More precisely, aOMP bases this decision on a threshold which is obtained using concentration of measure. The procedure stops when the current residual only consists of noise. By doing so, it prevents overfitting (i.e.\ to approximate noise) by using too many atoms and thus also substantially contributes to reducing computational times. As will be seen later, this feature of aOMP will be responsible for the faithful representation of different image features in the different quantitative parameter maps.\\
For more details, we refer the interested reader to \cite{pali2023dictionary} for the theoretical aspects and \cite{pali2021adaptive} for a previous application of the employed DL- and SC-algorithms within the context of qualitative image reconstruction.

\subsection{Notation}
We now introduce some abuse of notation which is required to simplify several expressions involving multiple dictionaries and sparse codes. For the parameter vector $\PP = [\PP_1, \ldots, \PP_P]^\trans$ we denote by $\Pd_j$  the operator which extracts  the $j$-patch from each component-vector of $\PP$, i.e. 
\begin{equation}\label{eq:patch_extraction_op}
\Pd_j \PP:= [\Pd_j \PP_1, \ldots, \Pd_j \PP_P]^\trans.
\end{equation}
Further, we abbreviate the approximation of the patch $\Pd_j \PP$ with a dictionary $\mathbf{\Psi}$ and a sparse code $\boldsymbol{\gamma}_j$ by 
\begin{eqnarray}\label{eq:dictionary_approx}
&\Pd_j \PP &\approx \mathbf{\Psi} \boldsymbol{\gamma}_j  \nonumber \\
\Leftrightarrow &[\Pd_j \PP_1, \ldots, \Pd_j \PP_P]^\trans &\approx [\mathbf{\Psi}_1 \boldsymbol{\gamma}_j^1, \ldots, \mathbf{\Psi}_P \boldsymbol{\gamma}_j^P]^\trans . \label{eq:sparse_approx}
\end{eqnarray}
From \eqref{eq:patch_extraction_op} and \eqref{eq:dictionary_approx}, we see that we intend to use the same patch-extraction operator on each vector component of $\PP$, but we intend to distinguish between $P$ different dictionaries and corresponding sparse codes for the approximation in \eqref{eq:sparse_approx}. This is important to note because the different vector components of $\PP$ might substantially differ in terms of local features and therefore, allowing the use of $P$ different dictionaries $\mathbf{\Psi}_p$, $p=1,\ldots, P$ might be preferable. Therefore, note that by $\mathbf{\Psi}$ we in general denote a set of $P$ different dictionaries (which potentially also might differ in the number of atoms $K$) and by $\boldsymbol{\gamma}_j$ we in general denote $P$ different sparse codes (also of potentially different dimension $K$). Also, by the relation $\|\boldsymbol{\gamma}_j \|_0 \leq S$ we intend 
\begin{equation}\label{eq:multi_sparsity}
\|\boldsymbol{\gamma}_j \|_0 \leq S \quad \Leftrightarrow \quad  \|\boldsymbol{\gamma}_j^1 \|_0 \leq S \, ,  \quad \ldots \, \, ,    \|\boldsymbol{\gamma}_j^P \|_0 \leq S,
\end{equation}
i.e.\ each component vector of $\PP$ is $S$-sparse with respect to its dictionary. Further, because of the use of the announced sparsity-adaptive sparse-coding algorithm aOMP, also the sparsity level of each sparse code $\boldsymbol{\gamma}_j^p$ might vary in \eqref{eq:multi_sparsity}.

\subsection{Proposed Adaptive Dictionary Learning-based Quantitative MR Image Reconstruction (ADLQMRI)}
For simplicity, we formally derive the proposed reconstruction scheme assuming a fixed dictionary. For the case where the dictionary is learned during the recnstruction as well, the proposed reconstruction method involves the solution of a problem of the type as described before in \eqref{DL_problem_L0}.\\
For a chosen sparsity level $S$ and a fixed dictionary $\mathbf{\Psi}$ the problem is formulated by
\begin{eqnarray}\label{eq:DL_reco_problem_qmri}
\underset{\PP, \{\boldsymbol{\gamma}_j\}_j}{\mathrm{min}} \frac{1}{2}\| \Wd^{1/2} (\Au  q_{\mathcal{T}} (\PP) -\YY ) \|_2^2 \nonumber \\   + \frac{\alpha}{2} \sum_j \| \Pd_j \PP - \mathbf{\Psi} \boldsymbol{\gamma}_j \|_2^2 \, \text{ s. t. } \,\| \boldsymbol{\gamma}_j\|_0 \leq S,
\end{eqnarray}
where $\Wd$ denotes a diagonal operator which contains the entries of a density compensation function and is used for pre-conditioning the problem in $k$-space \cite{pruessmann2001advances}.
Directly solving \eqref{eq:DL_reco_problem_qmri} is challenging. Mathematically, the problem is non-linear because of the signal model $q_{\mathcal{T}}$, non-convex because the constraint involving the $L_0$-norm and further possibly computationally  demanding because of the operator $\Au$. We therefore introduce auxiliary variables $\XX:=q_{\mathcal{T}} (\PP)$ (implicitly $T$ auxiliary variables $\XX_i = q_{t_i}(\PP)$ for $i=1, \ldots, T$)  as well as  $\UU:=  \PP$ and relax the equality constraints by including them in the form of quadratic penalty terms in our minimization problem, yielding
\begin{eqnarray}\label{splitting_DLQMRI_problem}
\begin{aligned}
    \underset{\XX, \PP, \UU, \{\boldsymbol{\gamma}_j\}_j }{\mathrm{min}}  \,  \frac{1}{2} \| \Wd^{1/2} ( \Au \XX - \YY )\|_2^2 +  \frac{\alpha}{2} \sum_j \| \Pd_j\UU - \mathbf{\Psi} \boldsymbol{\gamma}_j \|_2^2  \\ + \frac{\beta}{2}\| \XX - q_{\mathcal{T}} (\PP)\|_2^2 + \frac{\eta}{2}\|\UU -  \PP\|_2^2 \quad \text{ s. t. } \quad \| \boldsymbol{\gamma}_j\|_0 \leq S.
    \end{aligned}
\end{eqnarray}
We solve problem \eqref{splitting_DLQMRI_problem} by alternating minimization, i.e.\ by subsequently minimizing \eqref{splitting_DLQMRI_problem} with respect to one variable and keeping the others fixed. In the following, we derive the involved sub-problems which need to be solved during the reconstruction.\\
\noindent\textbf{Sub-Problem 1:} Assuming $\XX$ , $\UU$ and $\PP$ are fixed, updating $\{\boldsymbol{\gamma}_j\}_j$ corresponds to solving
\begin{equation}\label{sparse_coding}
    \underset{\{\boldsymbol{\gamma}_j\}_j }{\mathrm{min}}  \,    \frac{\alpha}{2} \sum_j \| \Pd_j\UU - \mathbf{\Psi} \boldsymbol{\gamma}_j \|_2^2    \quad \text{ s. t. } \quad \| \boldsymbol{\gamma}_j\|_0 \leq S.
\end{equation}
which is a sparse-coding problem and corresponds to finding the patch-wise sparse representation of $\UU$ with respect to the given dictionary $\mathbf{\Psi}$. For the case where the dictionary $\mathbf{\Psi}$ is jointly learned during the reconstruction, problem \eqref{sparse_coding} involves the minimization over the dictionary as well, yielding
\begin{equation}\label{eq:dico_learning_n_sparse_coding}
    \underset{\mathbf{\Psi} \in \mathcal{D}_{d,K}, \, \{\boldsymbol{\gamma}_j\}_j }{\mathrm{min}}  \,    \frac{\alpha}{2} \sum_j \| \Pd_j\UU - \mathbf{\Psi} \boldsymbol{\gamma}_j \|_2^2    \quad \text{ s. t. } \quad \| \boldsymbol{\gamma}_j\|_0 \leq S.
\end{equation}
  Note that although within the stated formulation, the sparsity level $S$ is fixed, since we use the sparsity- and dictionary-size adaptive aITKrM and aOMP algorithms, the set of admissible dictionaries $\mathcal{D}_{d,K}$ and the sparsity level $S$ are adaptively chosen depending on the set of patches $\{\Pd_j \UU\}_j$ and possibly vary among the different components of the vector $\PP$ as well as at each iteration. In particular, it is expected that $S$ tends to be chosen smaller at earlier iterations in order to reduce a large portion of noise and artefacts and larger at later iterations to be able to well-approximate the patches with high accuracy, see also  \cite{pali2021adaptive}.
  Also, note that for learning the dictionary, aITKrM uses iterative thresholding as sparse coding algorithm and $K$ residual means to update the dictionary \cite{schnass2018convergence}. As shown in \cite{pali2022average}, iterative thresholding instead of OMP for training the dictionary is considerably faster but at the same time competitive. Last, note that only for the final estimation of the sparse codes, i.e.\ after the dictionary has been learned, we use aOMP \cite{pali2021adaptive} for the sparse approximation of the patches.\\
\noindent\textbf{Sub-Problem 2:} Assuming $\XX$,  $\PP$ and $\{\boldsymbol{\gamma}_j\}_j$ are fixed, we update $\UU$ by solving 
\begin{equation}\label{sub_problem_u}
    \underset{\UU}{\mathrm{min}} \,  \frac{\alpha}{2} \sum_j \| \Pd_j\UU - \mathbf{\Psi} \boldsymbol{\gamma}_j \|_2^2  +  \frac{\eta}{2}\|\UU -  \PP\|_2^2.
\end{equation}
By taking the derivative with respect to $\UU$, setting it to zero an re-arranging, we obtain a linear system
\begin{equation}
     \big( \alpha \sum_j\Pd_j^\trans \Pd_j + \gamma\,\Id  \big) \UU  =   \alpha \sum_j\Pd_j^\trans \mathbf{\Psi} \boldsymbol{\gamma}_j + \gamma \PP.
\end{equation}
Then, assuming circular boundary conditions, strides of one for the patch-extraction operators  and by utilizing the identity $\sum_j\Pd_j^\trans \Pd_j = \alpha^{\prime}\,  \Id$, where $\alpha^{\prime}$ corresponds to a factor which comes from the overlapping of the pixels and which can be absorbed in the regularization parameter $\alpha$, we obtain $\UU$ as a closed-form solution
\begin{equation}\label{dl_approximation}
    \UU = \frac{1}{\alpha + \gamma} \big( \alpha  \sum_j\Pd_j^\trans \mathbf{\Psi} \boldsymbol{\gamma}_j + \gamma\, \PP \big).
\end{equation}
\noindent\textbf{Sub-Problem 3:} Updating the image $\XX$ assuming fixed $\UU$, $\PP$ and $\{\boldsymbol{\gamma}_j\}_j$ requires solving the problem
\begin{equation}\label{sub_problem_x}
\min_{\XX} \frac{1}{2} \| \Wd^{1/2} ( \Au \XX - \YY )\|_2^2 +  \frac{\beta}{2}\| \XX - q_{\mathcal{T}} (\PP)\|_2^2
\end{equation}
which is equivalent to solving a linear system $\Hd \XX = \mathbf{b}$ with
\begin{eqnarray}\label{lin_system}
\Hd &= &\Au^\herm \Wd \Au + \beta \, \Id \\
\mathbf{b} &= &\Au^\herm \Wd \YY + \beta\, q_{\mathcal{T}} (\PP),
\end{eqnarray}
using an iterative solver, e.g.\ a conjugate-gradient (CG) method. Note that because the operator $\Hd$ in \eqref{lin_system} has an approximately Toeplitz structure, there exist efficient implementations which exploit this structure and can make the application of the normal operator $\Au^\herm \Wd \Au$  faster by orders of magnitude compared to the separate application of $\Au^\herm$ and $\Au$, see \cite{wajer2001major} for more details.\\
\noindent\textbf{Sub-Problem 4:} Finally, updating $\PP$ assuming fixed $\XX$ , $\UU$ and $\{\boldsymbol{\gamma}_j\}_j$ involves solving the non-linear problem
\begin{equation}\label{sub_problem_p}
\min_{\PP} \frac{\beta}{2}\| \XX - q_{\mathcal{T}} (\PP)\|_2^2 + \frac{\eta}{2}\| \UU -  \PP\|_2^2
\end{equation}
by a non-linear optimization method, e.g.\ LBFGS-method \cite{liu1989limited} with bounds to ensure that the $\PP$ remains within physiologically realistic value ranges.\\
It is worth noting how the employed DL-based regularization acts as a regularizer for $\PP$. From \eqref{sub_problem_p}, we see that $q_{\mathcal{T}} (\PP)$ is on the one hand enforced to be close to $\XX$, where $\XX$ is coupled to $q_{\mathcal{T}} (\PP)$ (and thus implicitly to the measured data $\YY$) by the quadratic penalty term in \eqref{sub_problem_x} weighted by $\beta / 2$ and thus can be indirectly seen as the data-consistency term. On the other hand, $\PP$ is imposed to be close to $\UU$ which corresponds to a linear combination of the quantity $\PP$ and its sparse approximation from the sparse codes $\{\boldsymbol{\gamma}_j\}$.
Further, we note how due to the used splitting strategy, the repeated application of the operator $\Au  q_{\mathcal{T}}$ which would be required to directly solve \eqref{eq:DL_reco_problem_qmri} with respect to $\PP$ is avoided. 
Algorithm \ref{algo:qmri_algo} summarizes the described steps in a reconstruction algorithm.

\begin{algorithm}
    \caption{Proposed QMR Image Reconstruction Algorithm using adaptive Dictionary Learning and Sparse Coding}
    \label{algo:qmri_algo}
    
    \begin{algorithmic}
        \REQUIRE $k$-space data $\Yu$, \\ initializations $\PP^{(0)}:=[\mathbf{R}_1^{(0)}, \mathbf{M}_0^{(0)}, \mathbf{a}^{(0)}]^\trans, \XX^{(0)}$, regularization parameters $\alpha, \beta, \eta >0$, dictionary $\mathbf{\Psi}^{(0)}$  
        \STATE $k \gets 0$
        \WHILE{  $k \leq T$}
            \IF{ learn dictionary}
                \STATE $\mathbf{\Psi}^{(k+1)},  \{\boldsymbol{\gamma}_j\}_j^{(k+1)} \gets$ solve \eqref{eq:dico_learning_n_sparse_coding} (adaptive DL + SC)
            \ELSE
                \STATE  $\{\boldsymbol{\gamma}_j\}_j^{(k+1)} \gets$ solve \eqref{sparse_coding} (adaptive SC)
            \ENDIF
            
            \STATE  $\UU^{(k+1)} \gets$ solve \eqref{sub_problem_u} (closed-form solution)
            \STATE  $\XX^{(k+1)} \gets$ solve  \eqref{sub_problem_x} (CG algorithm)
            \STATE  $\PP^{(k+1)} \gets$ solve  \eqref{sub_problem_p} (LBFGS algorithm)
            \IF{ $\| \mathbf{R}_1^{(k+1)} - \mathbf{R}_1^{(k)} \|_2 / \| \mathbf{R}_1^{(k)}\|_2 < 10^{-3}$}
                \STATE Stop
                \STATE $\PP:=[\mathbf{R}_1^{(k+1)}, \mathbf{M}_0^{(k+1)}, \mathbf{a}^{(k+1)}]^\trans$
            \ENDIF
            \STATE $k \gets  k+1$
         \ENDWHILE
         \STATE $\PP:=[\mathbf{R}_1^{(T)}, \mathbf{M}_0^{(T)}, \mathbf{a}^{(T)}]^\trans$
    \end{algorithmic}
\end{algorithm}

\subsection{Initialization of the Reconstruction Algorithm}
A good starting point $\PP_0$ for the optimization can be obtained by applying a non-linear optimization algorithm for $\PP$ from the initial estimate $\XX_0:=\Au^{\sharp} \YY$, where $\Au^{\sharp}:=\Au^\herm \Wd$. For a chosen patch-dimensionality $d$, i.e.\ 2D patches of shape $\sqrt{d}\times \sqrt{d}=4 \times 4$  which were used in this work, we set $K=4d=64$ to initialize the dictionary $\mathbf{\Psi}_0 \in \mathcal{D}_{d,K}$ by randomly selecting $K$ patches from the initial quantitative maps $\PP_0$ and normalizing them with respect to the $\ell_2$-norm. Note again, that as visible from \eqref{eq:patch_extraction_op} and \eqref{eq:sparse_approx}, we are working with $P$ different dictionaries which are used for the corresponding quantitative parameters.

\section{Experiments}\label{sec:experiments}
Here, we apply the proposed method to a $T_1$-mapping QMRI reconstruction problem. First, we provide  information about the signal model $q_{\mathcal{T}}$ used in the experiments. Then, we describe the data acquisition as well as other methods used for comparison.

\subsection{The $\mathbf{R}_1$-Signal model}
The model involves the application of an  inversion pulse and the measurement of the $\mathbf{R}_1$-recovery of the magnetization. 
The unknown parameters for an entire slice are collected in the parameter vector $\PP$ and are given by $\mathbf{M}_0$, $\boldsymbol{a}$ and $\mathbf{R}_1$. For a time-point $t>0$, the signal model $q_t$ in \eqref{eq:q_t}  for this acquisition is given by
\begin{equation}\label{eq:signal_model}
    q_t(\mathbf{M}_0, \boldsymbol{a}, \boldsymbol{R}_1) = \mathbf{M}_0^* - (\mathbf{M}_0 + \mathbf{M}_0^*) \, \mathrm{exp}\{-t\,  \mathbf{R}_1^*\}
\end{equation}
where the the steady-state magnetization $\mathbf{M}_0^*$ is given by 
\begin{equation}\label{eq:steady_magn}
    \mathbf{M}_0^* = \frac{\mathbf{M}_0^* \cdot \mathbf{R}_1}{\mathbf{R}_1^*}
\end{equation}
and the apparent longitudinal relaxivitiy $\mathbf{R}_1^*$ is given by
\begin{equation}\label{eq:long_rel}
    \mathbf{R}_1^* = \mathbf{R}_1 - \frac{\ln(\cos \boldsymbol{a})}{T_R}.
\end{equation}
The time points $t$ for which MR data is acquired and the repetition time TR are determined by the MR acquisition parameters. Note that in \eqref{eq:signal_model}, \eqref{eq:steady_magn} and \eqref{eq:long_rel}, all operations are to be understood component-wise. 
The flip angle $\boldsymbol{a}$ is chosen depending on other MR sequence parameters. Nevertheless, due to the interaction between RF-excitation field and the low RF-wavelength within tissue at the used field strength of 7T, $\boldsymbol{a}$ can deviate from the chosen value and will spatially vary. This is especially challenging for higher magnetic fields used for example for 7\,T MR scanners \cite{ladd2018pros}.
In clinical practice,  the relaxation time $\mathbf{T}_1 := \mathbf{1} / \mathbf{R}_1$ is often used rather than the relaxivity parameter $\mathbf{R}_1$. 

\subsection{Data Acquisition}
In-vivo experiments were carried out on 10 healthy subjects from the NeuroMET cohort on a 7 Tesla whole-body Magnetom MR scanner (Siemens Healthineers, Erlangen, Germany) using a 1Tx/32Rx head coil (NOVA Medical, Wilmington, USA). We sequentially acquired 40 slices covering the brain with a spatial resolution of $1 \times 1 \times $2 mm$^3$ covering a field-of-view of $224 \times 224 \times 80$ mm$^3$. Data acquisition was carried out using a continuous Golden angle radial acquisition \cite{winkelmann2006optimal} after a single non-selective inversion pulse. For each slice, $N_{\theta} =  1504$ radial lines were acquired with TR$ =  7.3$ ms, echo time TE$ =   4.1$ ms and a flip angle $a = 5^\circ$. For the image reconstruction, the data of each slice was then split into $T=125$ time points, each with 12 radial lines. The total scan time of all 40 slices was 10 min.
In addition to this quantitative $\mathbf{T}_1$ mapping scan, an additional 3D MP2RAGE anatomical reference scan was also obtained (0.75 mm$^3$ isotropic resolution, TR$ = 7.2$ ms, TE$ = 2.5$ ms, total scan time = 12 min). 
The study was approved by the ethics committee of the Charité university hospital (EA2/121/19, 10.10.2019), and was conducted in accordance with the declaration of Helsinki.

\subsection{Numerical simulation}
In order to quantitatively compare the proposed ADLQMRI approach to other model-based QMRI approaches we carried out numerical simulations based on the BrainWeb data  \cite{cocosco1997brainweb}. Different $\mathbf{R}_1$- and $\mathbf{M}_0$-maps were created by assigning realistic values to the tissue segmentation provided by BrainWeb. For the flip angle $\boldsymbol{a}$, a Gaussian profile was simulated with a peak value of $a_{\mathrm{max}}=8^\circ$ in the center of the brain. MR data acquisition of 20 slices was simulated using the same parameters as for the in-vivo scans mentioned above. 

\subsection{Methods of Comparison}\label{subsec:methods_comparison}
Here, we briefly describe the used methods of comparison used in this work. 
\begin{itemize}
    \item ADLQMRI - the proposed DL- and SC-based regularization
    \item MAP - Model-Based Acceleration of Parameter Mapping \cite{Tran-Gia2013}, i.e.\ no regularization is imposed on $\PP$.
    \item TV -  total variation (TV) minimization based regularization\cite{Hamilton2017}
    \item Wl - Wavelet-based regularization \cite{daubechies1992ten}, \cite{Zhao2014}, \cite{Wang2019b}
    \item Sh -  Shearlet-based regularization \cite{guo2006sparse}, \cite{labate2005sparse}, \cite{Ma2017c}
    \item DL + Fit - denotes a DL- and SC-based method  for which first, qualitative images are reconstructed using DL and SC and a non-linear fit is subsequently applied to estimate $\PP$  \cite{Doneva2010}
\end{itemize}
\noindent \subsubsection{Model-based Acceleration of Parameter Mapping (MAP)}
In \cite{tran2013model}, it is proposed to estimate the quantitative parameters $\PP$ by first carrying out a GRAPPA operator gridding (GROG) interpolation \cite{seiberlich2007non} of the non-Cartesian $\YY$ to a Cartesian $k$-space $\tilde{\YY}$ for each receiver coil. By doing so, the operator $\Au$ can be replaced by a simple FFT-operator and the parameters $\PP$ are then iteratively estimated. First,  a zero-filled reconstruction $\XX_t$ for each interpolated radial line of $\tilde{\YY}$ is obtained using the IFFT-operator. Then, an estimation $\hat{\PP}$ of $\PP$ is obtained by fitting \eqref{eq:signal_model} to $\XX_t$ using a Levenberg-Marquardt (LM) algorithm \cite{more1978levenberg}. A fully sampled Cartesian $k$-space data  $\hat{\YY}_t$ for each $\XX_t$ is then calculated using $\hat{\YY}_t = \Au \, q_{\mathcal{T}} (\hat{\PP})$. Then, all data points which were originally acquired are replaced in $\hat{\YY}_t$ by the corresponding values of $\tilde{\YY}$. Finally, an updated version of $\XX_t$ is reconstructed and the iteration is repeated. 
\noindent \subsubsection{Sparsity-Based Methods}
For the TV-, Sh-, and Wl-based regularization, we formulate the reconstruction problems as 
\begin{equation}\label{eq:sparsity_methods}
\min_{\PP} \frac{1}{2} \| \Wd^{1/2} ( \Au \, q_{\mathcal{T}} (\PP) - \YY )  \|_2^2 + \alpha\, \| \mathbf{\Phi}\PP \|_1,
\end{equation}
where $\mathbf{\Phi}$ denotes the corresponding sparsifying transform, i.e.\ a 2D finite-differences operator for the TV-approach, a Haar-Wavelet basis \cite{stankovic2003haar} for Wl and a Shearlet-system \cite{Ma2017c} for Sh. Similar as in our approach, the solution of problem \eqref{eq:sparsity_methods} is approached by variable splitting. The different resulting sub-problems then require the use of a CG-method for solving for the qualitative images, of LBFGS for the non-linear sup-problem and soft-thresholding \cite{donoho1995noising} / iterative clipping \cite{chambolle2005total} for the sub-problem promoting the sparsity of the sought solution.

\noindent \subsubsection{DL+Fit}
To investigate the effect of the proposed splitting strategy, we also compared our approach to a different method which also uses DL and SC \cite{Doneva2010}. In contrast to our approach,  the method in \cite{Doneva2010} employs  DL- and SC-based regularization on the image data $\XX$ rather than on $\PP$. As a first step, qualitative images are reconstructed by imposing sparsity with respect to learned dictionaries as regularization. Then, in a second step, a non-linear fit is applied to estimate the vector $\PP$ from $\XX$, i.e.\
\begin{equation}\label{eq:DL_comparison_method}
\left \{
\begin{aligned}
&\underset{\PP}{\mathrm{min}}\, \frac{1}{2}\| \XX - q_{\mathcal{T}}(\PP)\|_2^2,\\
&\text{ where } (\XX, \{\boldsymbol{\gamma}_j\}_j) \in \underset{\tilde{\XX}, \{\boldsymbol{\tilde{\gamma}}_j\}_j}{\mathrm{argmin}} \frac{1}{2}\| \Wd^{1/2} ( \Au \tilde{\XX} -\YY ) \|_2^2 \\
&\qquad \quad \, \,+ \frac{\alpha}{2} \sum_j \| \Pd_j \tilde{\XX} - \mathbf{\Psi} \boldsymbol{\tilde{\gamma}}_j \|_2^2 \, \, \text{  s. t. }  \| \boldsymbol{\gamma}_j\|_0 \leq S
\end{aligned}
\right .
\end{equation}
From \eqref{eq:DL_comparison_method}, one can see that no regularization is explicitly imposed on the quantitative parameters $\PP$, but the regularization is implicitly encoded on the vector of images $\XX$ which is assumed to be patch-wise well-approximated by a sparse linear combination of the atoms of the dictionary $\mathbf{\Psi}$. For approximately imposing the constraint $\XX = q_{\mathcal{T}} (\PP) $, a LM fitting routine was used to minimize the squared error between the two vectors.
Because the sequence of qualitative images also exhibits correlation with respect to the direction, it seems a natural choice to use 3D patches instead of 2D patches as for our approach. The size of the 3D patches for DL+Fit was therefore chosen to be $4 \times 4 \times 6$, i.e.\ $d=96$ with $K=4d=384$.

\begin{figure*}[t]
\centering\includegraphics[width=\linewidth]{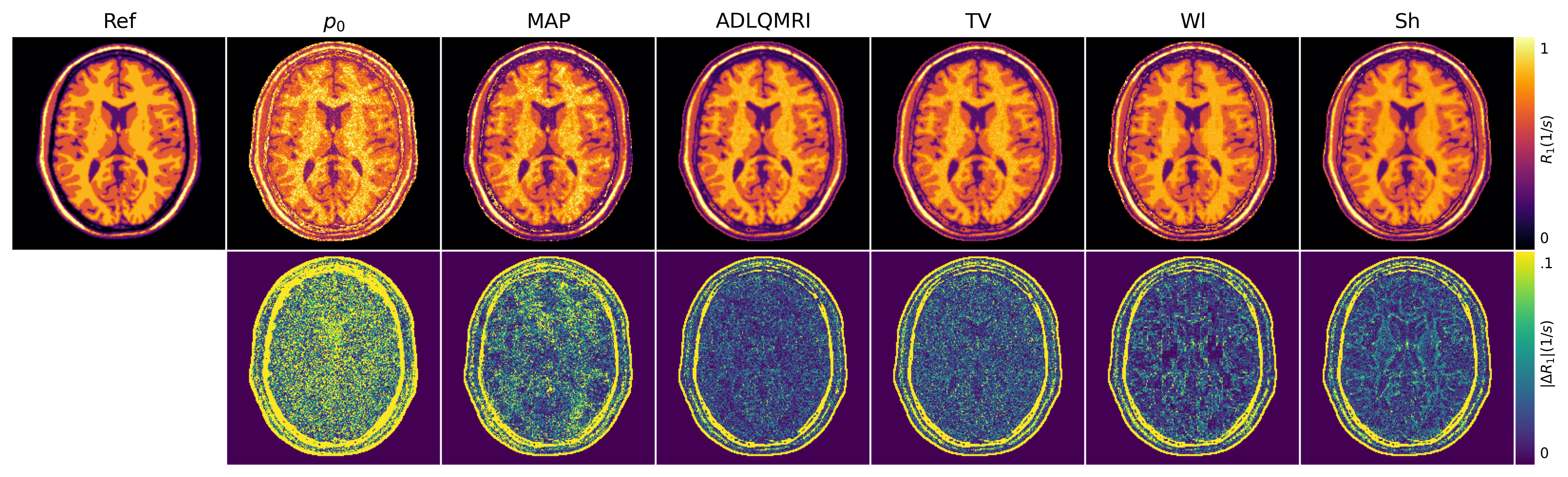}
\caption{Results of numerical simulation compared to the ground truth reference $\mathbf{R}_1$-map (Ref). $\mathbf{R}_1$-maps (top row) and absolute difference to the ground truth ($|\Delta R_1|$, bottom row) are shown for the initialisation ($p_0$) and different model-based image reconstruction approaches.}
\label{fig:xcat_reg_par_im}
\end{figure*}

\subsection{Parameter Reconstruction}
The acquired data was binned into dynamics each with radial lines. This reduces the number of times the model $q_{\mathcal{T}}$ has to be evaluated and hence reduces the reconstruction time. 
For the numerical simulations, the regularization parameters $\alpha$, $\beta$ and $\eta$ were selected for each reconstruction scheme separately based on the lowest root mean squared error (RMSE) for a central slice of the brain phantom.  This slice was then excluded from the subsequent  analysis.  For the in-vivo data, the regularization parameters were chosen using one slice of one subject based on visual inspection. The optimization was stopped if the relative change of the $\mathbf{R}_1$-map between two iterations was below $10^{-3}$. We based the decision on $\mathbf{R}_1$ since it is the clinically relevant parameter. The number of iterations for the CG-module for solving problem \eqref{lin_system} was set to five, which, despite being relatively small, suffices to achieve convergence because of the use of the operator $\Wd$ which is used for pre-conditioning the system.\\
In addition, we point out that the ranges of the values of quantitative parameters in general might differ from each other and thus different regularization parameters $\alpha_1, \ldots, \alpha_P$ could be used as well in the problem formulation (11). However, in our approach, we normalized the quantitative parameters such that one scalar regularization parameter $\alpha$ is sufficient. The normalization is taken into consideration in the non-linear signal model such that the application of $\Ad_I q_{\mathcal{T}}$ matches the range of the acquired $k$-space data. 

\subsection{Evaluation}
All methods were evaluated in terms of RMSE as well as peak signal-to-noise ratio (PSNR) which for $\mathbf{M}_0$, $\boldsymbol{a}$ and $\mathbf{R}_1$ were averaged over over all slices. Further, the standard deviations of the corresponding RMSE and PSNR are reported as measures of stability of the methods. Note that we abstain from reporting image similarity-based measures like the structural similarity index measure (SSIM) \cite{wang2004image} because the images we reconstruct are quantitative parameters maps whose values have a physiological meaning.

\section{Results}\label{sec:results}

\subsection{Numerical Simulation}
\begin{figure}[!h]
\centering\includegraphics[width=\linewidth]{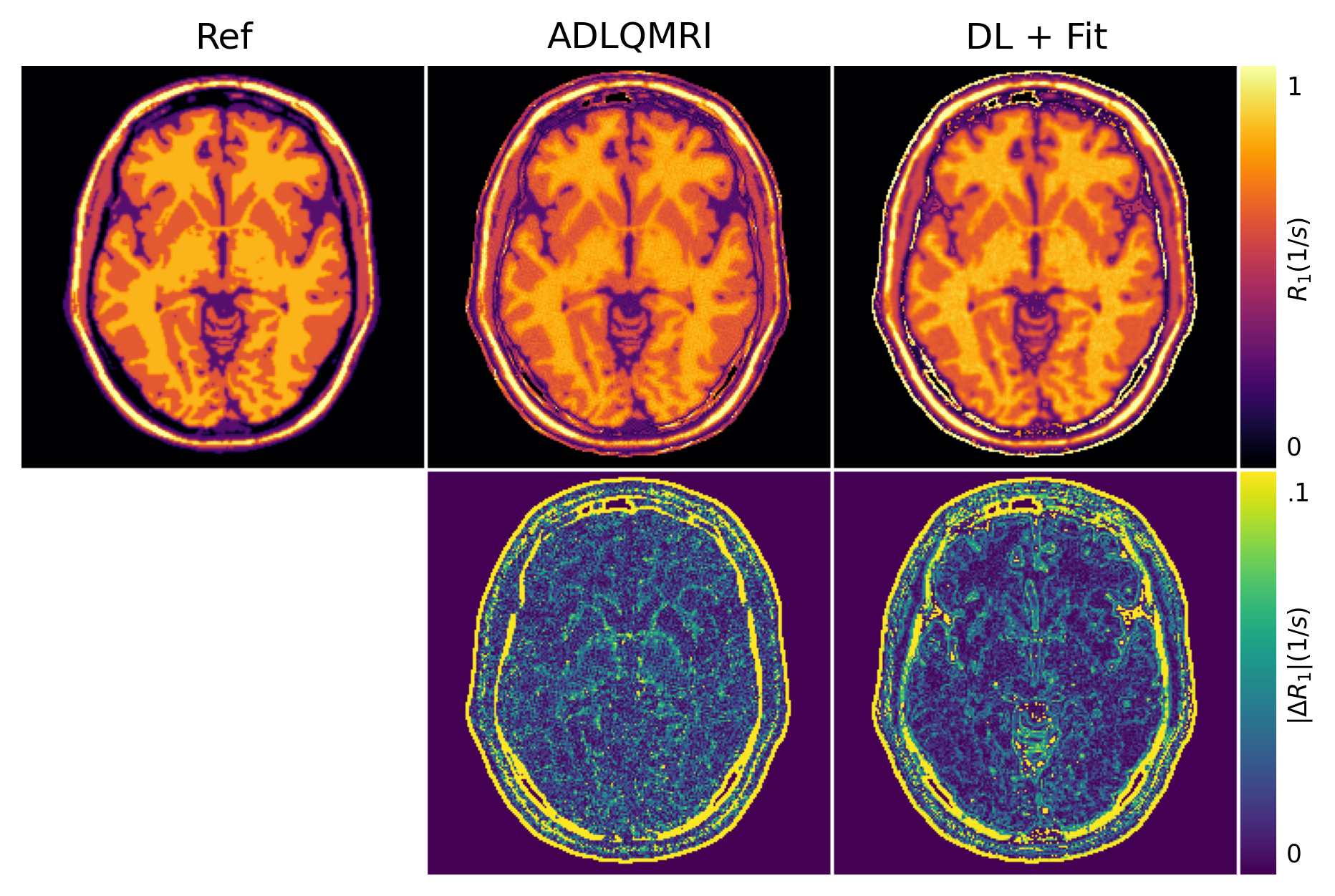}
\caption{Comparison of the proposed ADLQMRI method to DL+Fit which first used DL and SC to reconstruct qualitative images and subsequently applies a non-linear data fit.}
\label{fig:xcat_dl_im}
\end{figure}

Figure \ref{fig:xcat_reg_par_im} shows the results of the numerical simulation. The starting point $\PP_0$ is strongly impaired by noise which is partially reduced by MAP. The regularized model-based reconstructions further improve the estimation of $\mathbf{R}_1$. The proposed ADLQMRI method leads to the most accurate $\mathbf{R}_1$ estimation which is also confirmed by Tables \ref{tab:rmse_results} and \ref{tab:psnr_results}. As can be seen from Figure \ref{fig:xcat_dl_im}, DL + Fit also leads to accurate parameter estimation  but requires much longer reconstruction times, see Table \ref{tab:recon_times}. Further, for all three parameters, the standard deviation for DL + Fit is considerably higher than for the proposed ADLQMRI. 

\begin{table}[h]
\centering
\caption{RMSE for the numerical simulations.}
\label{tab:rmse_results}
\setlength{\tabcolsep}{3pt}
\begin{tabular}{l c c c} \hline
 & $\mathbf{M}_0$ & $\mathbf{a} (^\circ)$ & $\mathbf{R}_1 (1/s)$  \\ 
 \hline
MAP &
 $0.162 \pm 0.018$ &
 $0.160 \pm 0.014$ &
 $0.124 \pm 0.031$ 
 \vspace*{0.0mm} \\
ADLQMRI &
 $0.039 \pm 0.003$ &
 $\mathbf{0.019 \pm 0.004}$ &
 $\mathbf{0.039 \pm 0.005}$ 
 \vspace*{0.0mm} \\
TV &
 $0.052 \pm 0.005$ &
 $0.046 \pm 0.010$ &
 $0.048 \pm 0.006$ 
 \vspace*{0.0mm} \\
Wl &
 $0.053 \pm 0.004$ &
 $0.054 \pm 0.006$ &
 $0.047 \pm 0.010$ 
 \vspace*{0.0mm} \\
Sh &
 $0.040 \pm 0.010$ &
 $0.041 \pm 0.013$ &
 $0.045 \pm 0.003$ 
 \vspace*{0.0mm} \\
DL+Fit &
 $\mathbf{0.037 \pm 0.010}$ &
 $0.055 \pm 0.021$ &
 $0.054 \pm 0.018$ 
 \vspace*{0.0mm} \\
 \hline 
 \end{tabular}
 \end{table}

\begin{table}[h]
\centering
\caption{PSNR for the numerical simulations.}
\label{tab:psnr_results}
\setlength{\tabcolsep}{3pt}
\begin{tabular}{l c c c} \hline
 & $\mathbf{M}_0$ & $\mathbf{a} (^\circ)$ & $\mathbf{R}_1 (1/s)$ \\ 
 \hline
MAP &
 $23.17 \pm 1.78$ &
 $23.84 \pm 1.15$ &
 $31.55 \pm 1.58$
 \vspace*{0.0mm} \\
ADLQMRI &
 $35.40 \pm 1.50$ &
 $\mathbf{42.62 \pm 2.25}$ &
 $\mathbf{41.55 \pm 1.34}$
 \vspace*{0.0mm} \\
TV &
 $33.08 \pm 1.77$ &
 $34.94 \pm 2.73$ &
 $39.73 \pm 1.43$
 \vspace*{0.0mm} \\
Wl &
 $32.79 \pm 1.45$ &
 $33.39 \pm 1.79$ &
 $40.00 \pm 0.83$
 \vspace*{0.0mm} \\
Sh &
 $35.35 \pm 1.39$ &
 $36.02 \pm 1.75$ &
 $40.21 \pm 1.08$
 \vspace*{0.0mm} \\
DL+Fit &
 $\mathbf{36.09 \pm 1.86}$ &
 $33.67 \pm 2.96$ &
 $39.02 \pm 1.43$
 \vspace*{0.0mm} \\
 \hline 
 \end{tabular}
 \end{table}

\begin{table}
\centering
\caption{Reconstruction time per slice}
\label{tab:recon_times}
\setlength{\tabcolsep}{3pt}
\begin{tabular}{l |r c r}
 & \multicolumn{3}{c}{Time/Slice (min)}\\ 
 \hline
$p_0$ &  $0.7 $ & $\pm$ & $0.1$
 \vspace*{0.0mm} \\
MAP &  $23.2 $ & $\pm$ & $4.6$
 \vspace*{0.0mm} \\
ADLQMRI &  $76.3 $ & $\pm$ & $25.9$
 \vspace*{0.0mm} \\
TV &  $3.9 $ & $\pm$ & $0.8$
 \vspace*{0.0mm} \\
Wl &  $5.7 $ & $\pm$ & $1.6$
 \vspace*{0.0mm} \\
Sh &  $9.0 $ & $\pm$ & $2.5$
 \vspace*{0.0mm} \\
DL+Fit &  $535.1 $ & $\pm$ & $31.3$\\
 \hline 
 \end{tabular}
 \end{table}

\subsection{In-Vivo Experiments}
\begin{figure}[t]
\centering\includegraphics[width=\linewidth]{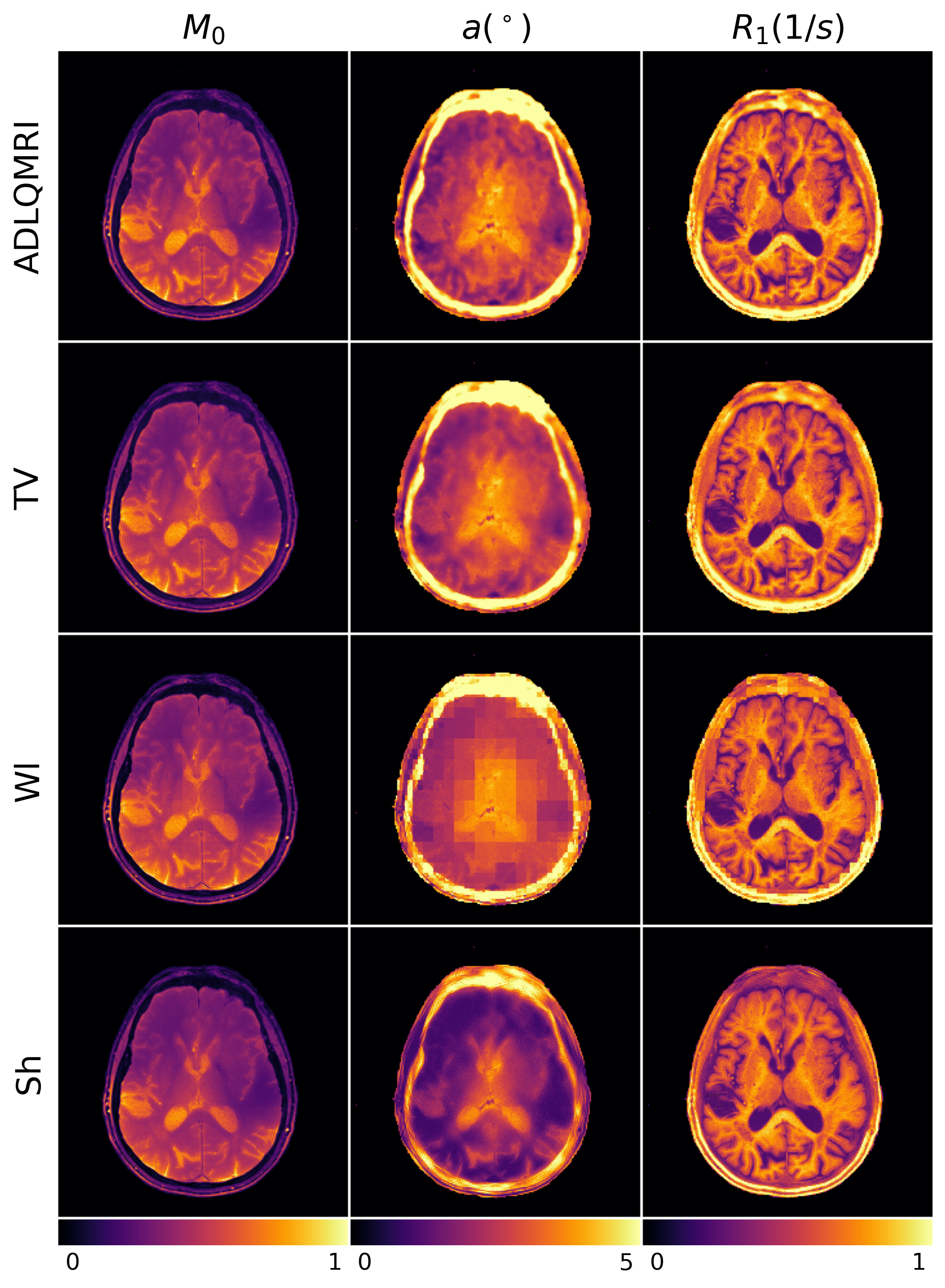}
\caption{In-vivo results of all three components of $\PP$ ($\mathbf{M}_0$, $\boldsymbol{a}$ and $\mathbf{R}_1$) for different model-based image reconstruction approaches. }
\label{fig:in_vivo_3par}
\end{figure}

\begin{figure}[!h]
\centering\includegraphics[width=\linewidth]{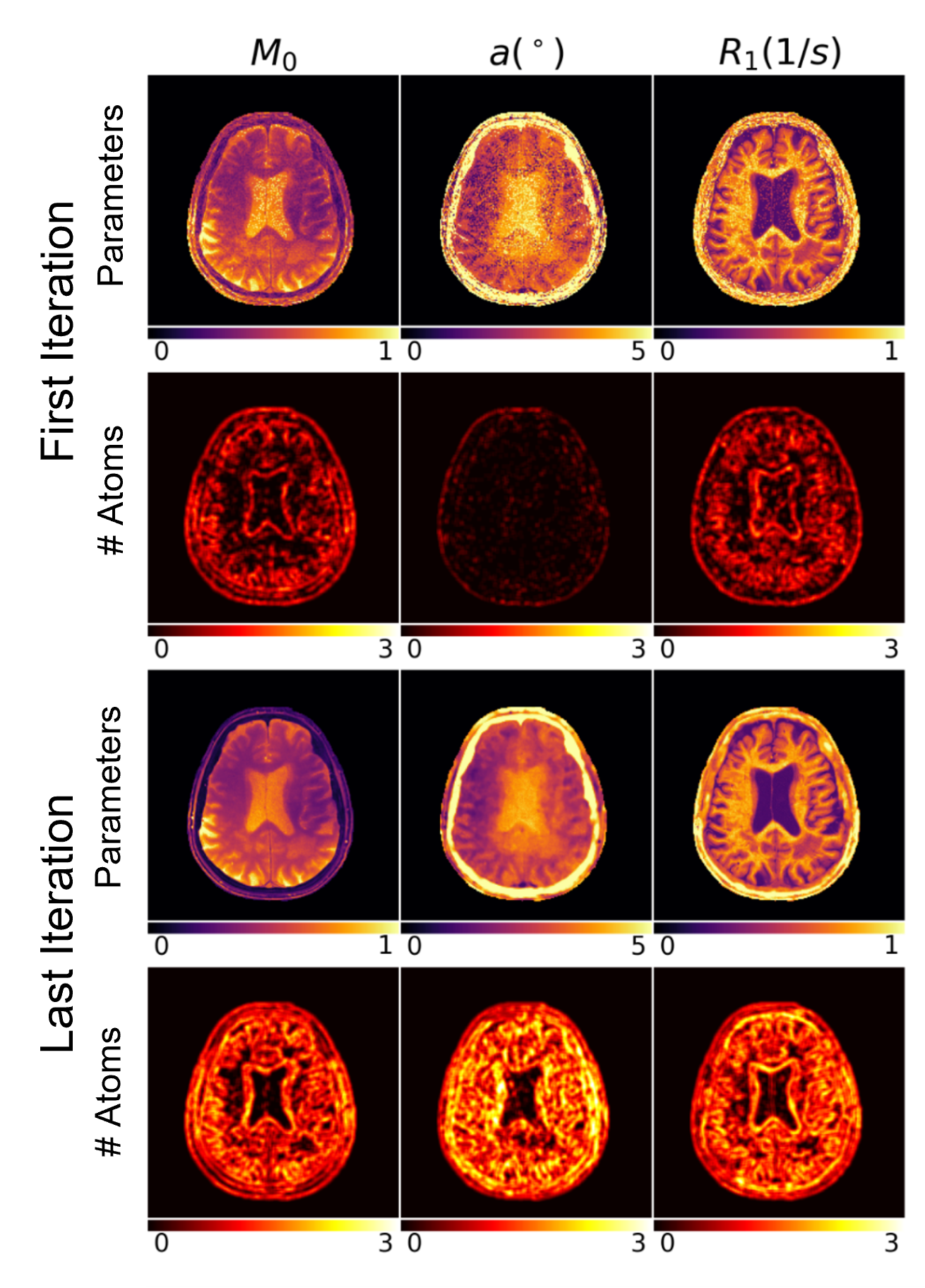}
\caption{Three components of $\PP$ ($\mathbf{M}_0$, $\boldsymbol{a}$ and $\mathbf{R}_1$) and the corresponding estimated sparsity-level for the first and final iteration of the proposed ADLQMRI method shown for a subject.}
\label{fig:in_vivo_sparse_codes}
\end{figure}

\begin{figure*}[!h]
\centering\includegraphics[width=\linewidth]{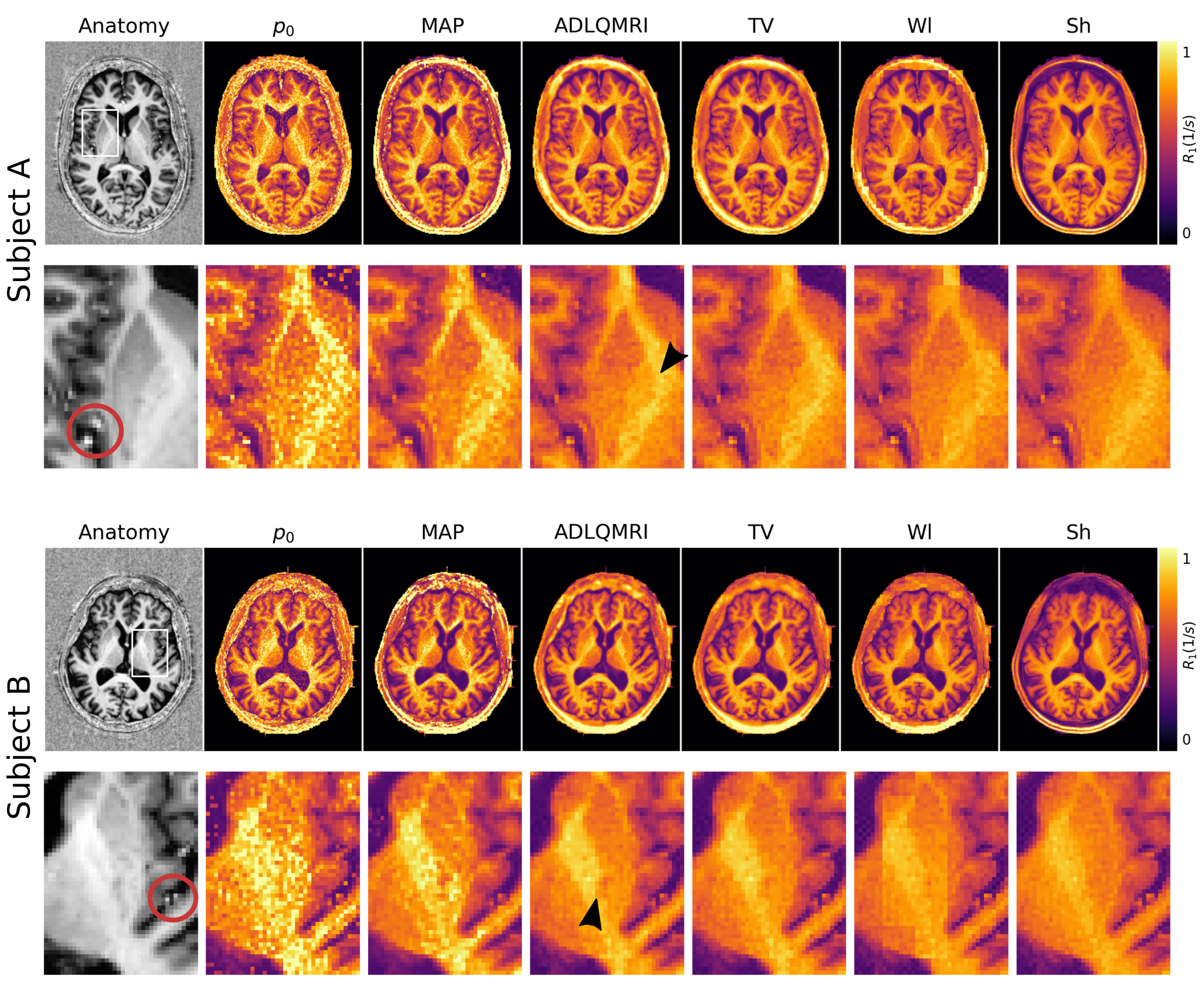}
\caption{In-vivo results showing $\mathbf{R}_1$ for different model-based image reconstruction approaches for two subjects. A $T_1$-weighted image reformatted to the same slice position as the $\mathbf{R}_1$ maps is shown as anatomical reference. The location of the zoomed image area is highlighted with a white rectangle in the anatomical reference. Black arrows highlight small anatomical features such as the Genu of the internal capsule near the Globus Pallidus (subject A) and the crossing region between Globus Pallidus and Putamen (subject B). Blood vessels are highlighted by red circles.}
\label{fig:in_vivo_patients}
\end{figure*}

\begin{figure*}[!h]
\centering\includegraphics[width=\linewidth]{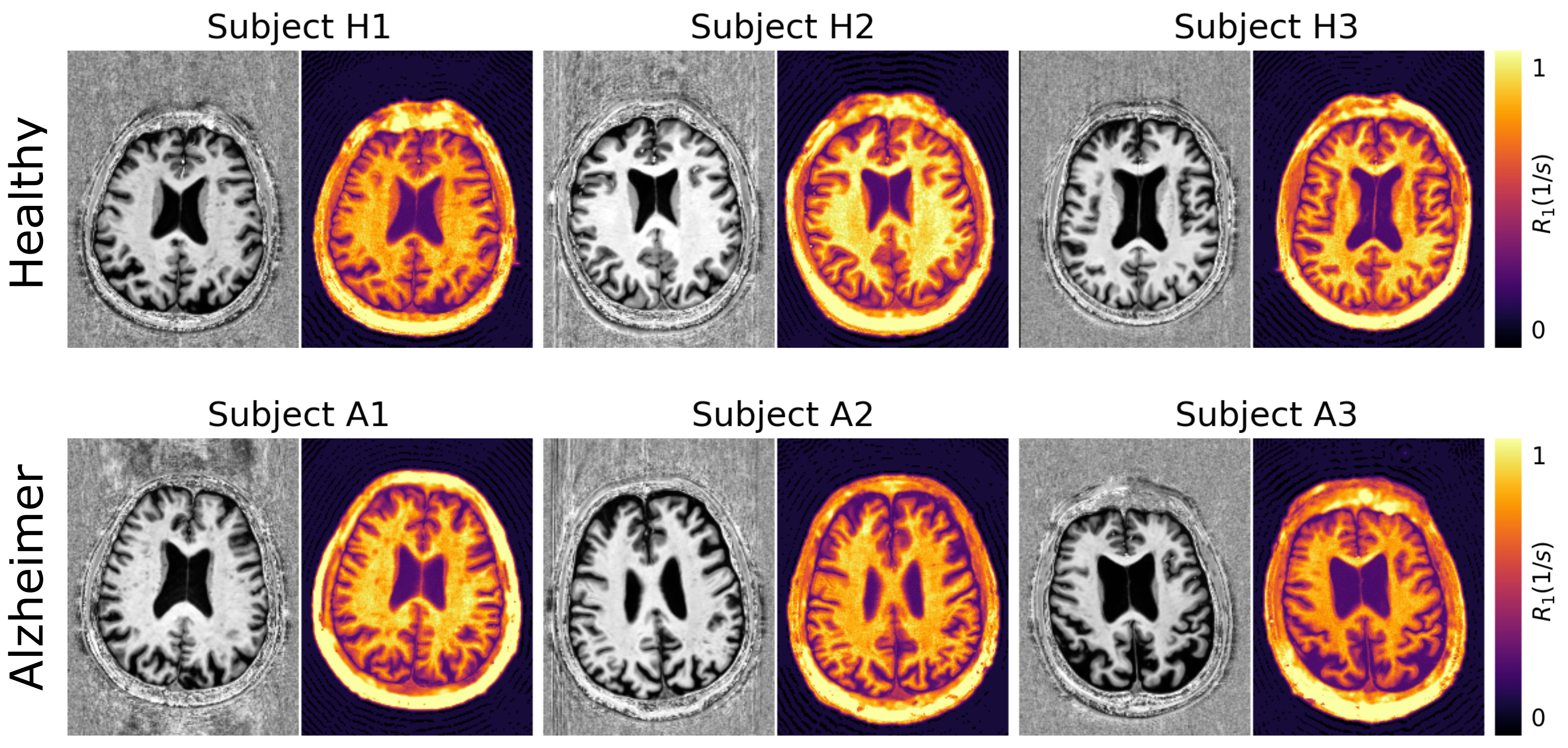}
\caption{A visual comparison of three healthy subjects (H1, H2 and H3) and three patients with known Alzheimer's disease (A1, A2 and A3) obtained with the proposed method ADLQMRI. The obtained $R_1$-maps are in good agreement with the intermediate qualitative images in terms of local features, both for the healthy subjects as well as for the patients. }
\label{fig:in_vivo_patients_vs_subjects}
\end{figure*}

Figure \ref{fig:in_vivo_3par} shows the results of all three parameters $\mathbf{M}_0$, $\boldsymbol{a}$ and $\mathbf{R}_1$ for a subject. The obtained results for $\mathbf{M}_0$ and $\mathbf{R}_1$ are comparable between all four methods. The largest differences can be seen for the flip angle $\boldsymbol{a}$. This is also in agreement with the numerical simulations, where the largest differences between ADLQMRI and the other model-based reconstruction methods could be seen for $\boldsymbol{a}$. For Wl, we see "patchy" artefacts appearing for $\boldsymbol{a}$, while Sh tends to deteriorate the contrast.
Further, to demonstrate the effectiveness of the employed DL- and SC-algorithms aITKrM and aOMP, in Figure \ref{fig:in_vivo_sparse_codes}, the average number of atoms used per patch is depicted. In the first and the third row of Figure \ref{fig:in_vivo_sparse_codes}, the estimates of the three parameters $\mathbf{M}_0$, $\boldsymbol{a}$ and $\mathbf{R}_1$ at the first and final iteration are shown, respectively. We clearly see how the noise present in the initial estimates is highly reduced from the first to the last iteration. Further, we see that the noise level in the initial estimates varies across the different components of the vector $\PP$.
In addition, the second and the third row of Figure \ref{fig:in_vivo_sparse_codes} show the patch-wise estimated sparsity level at the first and the last iteration, respectively. Here, we first see that the highest number of used atoms for the sparse approximation amounts at most  to be three for all parameters $\mathbf{M}_0$, $\boldsymbol{a}$ and $\mathbf{R}_1$. Further, we see that the number of required atoms varies across the parameters as well as across the local position of the image depending on the content which needs to be represented. Smoother image regions only require a small number of atoms, while regions with edges and diagnostic details require a more precise representation using a larger number of dictionary atoms. Last, we see that the number of atoms required for the sparse representation also varies across the iterations, tending to be lower at the first iteration and higher at the last.\\ 
The in-vivo results in Figure \ref{fig:in_vivo_patients} compare the obtained $\mathbf{R}_1$ maps for two subjects to a reformatted $T_1$-weighted anatomical scan. ADLQMRI leads to successful noise suppression while preserving small details. TV leads to comparable image quality whereas Wl and Sh lead to block-artefacts and loss of small details. Further, Figure \ref{fig:in_vivo_patients_vs_subjects} shows a comparison of the $R_1$-maps of three healthy subjects (H1, H2 and H3) and three patients with known Alzheimer's disease (A1, A2, A3).

\section{Discussion}\label{sec:discussion}
In the following, we discuss different aspects of the proposed reconstruction method in more detail, highlight differences and similarities to other works, discuss the limitations and give an outline for possible future work.

\subsection{The Employed Splitting and Regularization Strategy}\label{subsec:splitting_n_reg}
\textcolor{black}{
The two main ingredients of the proposed reconstruction method are on the one hand to decouple the measurement operator $\Au$ from the signal model $q_{\mathcal{T}}$ and, on the other hand, to impose the regularization directly on the quantitative parameters of interest rather than on the qualitative images. These two steps serve two different purposes. First, the decoupling of $\Au$ and $q_{\mathcal{T}}$ avoids the repeated application of the (possibly) computationally expensive operator $\Au$ within a non-linear optimization routine as LBFGS or LM. Second, the rationale of employing the regularization directly on the parameter-maps $\PP$  serves as a considerable dimensionality reduction. The computational time required by the sparse-coding step is highly reduced because the sparse approximation of the patches $-$ which is the most time-consuming component of any reconstruction algorithm based on patch-wise sparsity with respect to a dictionary $-$ has to be performed only for the components of $\PP$ rather than for all qualitative images $\XX_1,\ldots, \XX_T$.  As can bee seen from Table \ref{tab:recon_times}, the comparison between DL+Fit \cite{Doneva2010} and the proposed ADLQMRI method shows a significant reduction in terms of reconstruction times by a factor of approximately seven ($\approx$ 1\,h 16\,min vs.\ 7\,h 45\, min). Further, the approach avoids noise- or errors-amplification due to the non-linearity of the model $q_{\mathcal{T}}$.  As can be seen from Tables \ref{tab:rmse_results} and \ref{tab:psnr_results}, the proposed method indeed outperforms DL + Fit by a large margin with respect to $\boldsymbol{a}$ and $\mathbf{R}_1$ while yielding comparable results for $\mathbf{M}_0$ in terms of the obtained average measure. Further, the standard deviation of the PSNR and the RMSE is highly reduced for all quantitative parameters.
}

\subsection{Adaptive Dictionary Learning and Sparse Coding}
\textcolor{black}{
Another noteworthy component of the proposed reconstruction algorithm is the use of the adaptive DL and SC algorithms aITKrM \cite{pali2018dictionary} and aOMP \cite{pali2021adaptive}. Note that in the literature, some authors also use the term "adaptive dictionary learning", see e.g.\ \cite{caballero2014dictionary}. However, the term in that context is related to so-called "blind Compressed Sensing" approaches where the sparsifying transforms, in our case the dictionary, are learned during the reconstruction. In our work, the concept of adaptivity stems from the work in \cite{pali2018dictionary} and intends the fact that the sparsity level $S$ and the overall number of atoms of the dictionary $K$ are jointly estimated during the reconstruction. This means that each iteration of the reconstruction algorithm, a  dictionary with potentially different size and sparsity level is learned. Therefore, the employed DL and SC algorithms are adaptive in both  senses. The dictionary-size and sparsity level-adaptivity have in fact a large impact which can be best seen from Figure \ref{fig:in_vivo_sparse_codes}. There, we see that using algorithms as for example $K$-SVD and OMP, which require an a-priori and global choice of $K$ and $S$ cannot be optimal. Clearly, the three parameters $\mathbf{M}_0$, $\boldsymbol{a}$ and $\mathbf{R}_1$ exhibit features and noise levels which vary across the three parameters, across the iterations as well as across the location of the image. Therefore, in addition to being optimally chosen, the adaptive choice of $S$ and $K$ provided by using aITKrM and aOMP as well reduces the number of required experiments for hyper-parameter tuning. In addition, because $S$ is never overestimated, as reported in \cite{pali2018dictionary}, it can accelerate the reconstruction while maintaining a similar performance as $K$-SVD and OMP.  
}
\subsection{Limitations}
\textcolor{black}{
At the current stage, the main limitation of the proposed approach is the (although highly reduced) overall long reconstruction time which can be attributed to the sparse-coding of the employed reconstruction regularization method  based on DL. Note however, that as previously mentioned in Subsection \ref{subsec:splitting_n_reg}, the proposed splitting strategy is explicitly designed to be well-suited for regularization methods which are known to be time-consuming, such as blind Compressed Sensing methods. 
Further, although the proposed method shows satisfactory reconstruction results, the involved splitting of the original reconstruction in sub-problems requires the choice of additional regularization parameters $\lambda, \beta$ and $\eta$, which $-$ up to this point $-$ can only be empirically chosen. However, we report that we have found the method to be relatively stable int terms of RMSE and PSNR with respect to changes in the regularization parameters.\\
In addition, the strategy used to decouple problem \eqref{eq:DL_reco_problem_qmri} into a series of simpler sub-problems which are subsequently solved in an alternating fashion raises the question about the convergence of the proposed Algorithm \ref{algo:qmri_algo} to a solution of \eqref{eq:DL_reco_problem_qmri}, also given its dependence on the regularization parameters $\lambda, \beta$ and $\eta$. As alternatives, one could instead consider non-linear extensions of the primal dual hybrid gradient (PDHG) method \cite{chambolle2011first} or the alternating direction method of multipliers (ADMM) \cite{boyd2011distributed} proposed in \cite{valkonen2014primal} and \cite{benning2016preconditioned}, respectively, whose convergence can be guaranteed under some technical problem-dependent conditions.\\
The presented approach's superiority was quantitatively evaluated merely on retrospectively simulated data. The lack of ground-truth quantitative parameter maps for in-vivo applications limits the possibility to verify the accuracy of the presented $T_1$ maps. However, the qualitative in-vivo $T_1$-weighted images give an indication of image features which should also be present in the reconstructed $T_1$-maps. Future studies are needed to verify the accuracy in-vivo in larger patient studies.
Further, note that Alzheimer’s disease leads to global changes of the brain structure and not to
localized changes as e.g. in the case of tumors. Thus, a differentiation between healthy subjects and patients cannot be based on the visual inspection of pathological features, as for example from Figure \ref{fig:in_vivo_patients_vs_subjects}, but requires a quantitative comparison of the $T_1$-maps on a large cohort of patients and subjects for which further clinical trials are needed. 
}

\subsection{Differences and Similarities to Previous Works}
\textcolor{black}{
Methodologically speaking, our method is strongly related to the one presented in \cite{Doneva2010} with a few essential differences. In \cite{Doneva2010}, hard data-consistency is enforced by estimating the missing $k$-space coefficients from the ones obtained by applying the Fourier-transform on the image which is assembled by the sparsely approximate patches. However, as can be seen from \eqref{eq:DL_comparison_method},  this step can only be motivated for a single-coil Cartesian acquisition but not for a non-uniform Fourier encoding operator as the one used in this work. Therefore, in our adaptation of the method of \cite{Doneva2010}, the hard data-consistency step is replaced by solving the sub-problem of \eqref{eq:DL_comparison_method} with respect to $\XX$.\\
Further, as already mentioned in Subsection \ref{subsec:methods_comparison}, the chosen splitting strategies  are different and are the reason to which we can attribute the  differences in terms of performance (see Table \eqref{tab:rmse_results}) and, most importantly, in terms of computational time (see Table \eqref{tab:rmse_results}).
}

\subsection{Outline and Future Work}
\textcolor{black}{
Although in this work we have used DL and SC as the regularization method of choice, the proposed splitting strategy is not limited to be applicable with DL and SC. In fact, a large variety of other learning-based methods, such as patch-wise analysis operator learning methods \cite{ravishankar2015efficient},  convolutional DL \cite{quan2016compressed},  \cite{chun2017convolutional}  and convolutional analysis operator learning \cite{Chun2020ConvolutionalAO}, or  reconstruction-adaptive neural networks (NNs)-based methods \cite{ulyanov2018deep}, \cite{yoo2021time} could be considered as well.
In addition, because the employed adaptive DL and SC algorithms aITKrM and aOMP well-adapt to the different contrasts of the parameter vector and reduce the computational time compared to $K$-SVD \cite{aharon2006k} and OMP \cite{pati1993orthogonal}, they could be as well applied to approaches using coupled DL \cite{song2019coupled}, where the number of times DL- and SC-algorithms have to be run is increased by the number of dictionary and sparse codes which are coupled. 
In addition, as the parameter images might as well share common local features, it might be desirable to also exploit this similarity in a similar fashion as investigated in \cite{song2019coupled}. There, image-patches with different contrasts are represented by a linear combination of different sparse representation which involve both contrast-specific dictionaries as well as a common sparse code to describe the the similarity and discrepancy between two different contrasts. Interestingly, the authors explicitly state that their approach could be extended for the case where the dictionaries used with the corresponding contrast-specific sparse codes, have a different number of atoms. Thus, we believe that a combination of the approach in \cite{song2019coupled} with the employed size-adaptive aITKrM for the task of quantitative image reconstruction using our proposed reconstruction scheme could yield promising results.\\
Last, we mention that the proposed approach could be utilized to obtain target-data for the development of data-driven methods based on supervised learning, e.g.\ Deep Learning.}

\section{Conclusion}\label{sec:conclusion}
\textcolor{black}{
In this work, we have proposed a method for the reconstruction of quantitative parameter-maps using dictionary learning (DL) and sparse coding (SC). By directly imposing the regularization on the different components of the sought quantitative parameter-maps, the time required for solving the resulting reconstruction problem is reduced by approximately a factor of seven compared to the approach where the sparsity with respect to the dictionary is imposed on the intermediate qualitative images. Further, because the different quantitative parameter-maps differ in terms of contrast, noise level and local features, we used adaptive DL- and SC-algorithms in which the total number of dictionary atoms as well as the optimal number of atoms to be used during the SC stage are adaptively chosen at each iteration and for each parameter-map separately. We have seen that the employed adaptive DL- and SC-algorithms well-adapt to the considered data and faithfully represent the images, allowing for an efficient patient-adaptive regularization method. Although the proposed method was applied to a $T_1$-mapping example in the brain, we expect it to be broadly applicable for other signal-models and organs as well. Further, the proposed problem formulation can as well be used with other possibly time-consuming regularization methods such as convolutional dictionary learning or analysis operator learning. 
}

\section*{Acknowledgments}
\textcolor{black}{
The results presented here have been developed in the
framework of the 18HLT05 QUIERO Project and 18HLT09 NeuroMET2 Project. These projects have received
funding from the EMPIR programme co-financed by the
participating states and from the European Union’s Horizon
2020 research and innovation program.\\
We further thank Dr.\ Marie-Christine Pali for the fruitful discussions about adaptive dictionary learning and sparse coding.
}
\bibliographystyle{IEEEtran}
\textcolor{black}{
\bibliography{IEEEabrv,references}
}
\end{document}